\documentstyle[11pt,russian]{article}
\begin{document}
\begin{center}{\bf TOPOLOGICAL INTERPRETATION OF DIRAC EQUATION AND 
GEOMETRIZATION OF PHYSICAL INTERACTIONS}\end{center}
\begin{center}O.A.OLKHOV\end{center}
\begin{center}{\it Semenov Institute of Chemical Physics, Russian Academy of 
Sciences, Moscow}\end{center}
\begin{center}{\it e-mail: olkhov@center.chph.ras.ru}\end{center}
\par\bigskip

\noindent     Earlier we have shown that interacting electron-positron and 
electromagnetic fields can be considered as a certain microscopic distortion of
pseudo-Euclidean properties of the Minkovsky 4-space-time. The known Dirac and 
Maxwell equations prove to be group-theoretical relations describing this
distortion (nonmetrized closed 4-manifold). Here we apply the above geometrical 
approach to obtain equations for a neutrino interacting with its weak field. 
These equations contain some new terms and demonstrate geometrical
mechanisms of gauge-invariance and P-T violation. Equations are also proposed 
for gravitational field and its microscopic quantum sources.
\bigskip

\noindent{\bf Introduction}
\par\medskip
There were many attempts to find out new interpretation of the mathematical 
formalism of quantum mechanics and geometrical representation of physical 
fields. Both these problems are usually considered separetly. In our 
investigation, these problems are combined within a unique hipothesis. It
was suggested that the relativistic quantum equations for the microscopic 
field sources and equations for these fields themselves describe some unified 
specific geometrical object---microscopic nonmetrized closed space-time 
4-manifold. 

In our previous
publications, we proposed this topological interpretation for the Dirac and 
Maxwell equations which describe interacting electron-positron and 
electromagnetic fields. These equations can be written in the form [1]
$$i\gamma_1\left (\frac{\partial}{\partial x_1}-ieA_1 \right)\psi-
\sum_{\alpha =2}^4 i\gamma_\alpha \left(\frac
{\partial}{\partial x_\alpha}-ieA_\alpha \right)\psi=m_e\psi,\eqno (1)$$
$$F_{kl}=\frac{\partial A_k}{\partial x_l}-\frac{\partial A_l}{\partial x_k},
\eqno (2)$$
$$\sum_{i=1}^4 \frac{\partial F_{ik}}{\partial x_i}=j_k^e,\quad
j_k^e=e\psi*\gamma_1\gamma_k\psi.\eqno (3)$$ 
Here $\hbar=c=1,\quad x_1=t,\quad x_2=x,\quad x_3=y,\quad x_4=z,\quad F_{kl}$ 
is the electromagnetic field tensor, $A_k$ is 
the 4-potential, $\gamma_k$ are the Dirac matrices, $\psi$ is the 
Dirac bispinor, and $m_e$ and $e$ are electron mass and charge, respectively.

It was shown that Eqs.(1-3) can be interpreted as group-theoretical 
relations that account for the topological and metric properties of some
unified microscopic nonmetrized closed space-time 4-manifold [2-5]. 
Equations (1-3)
describe the above properties in this manifold-covering space and, this space
proves to be a specific space--- a space with so-called "semimetrical 
parallel translation" [6]. In this space,
$$\bigtriangledown_lG_{mn}=-\Gamma_lG_{mn},$$
where $\bigtriangledown_l$ is a covariant derivative, $G_{mn}$ is a metric and
$\Gamma_l$ is a 4-vector (note that for the Riemann space in general relativity
this derivative is zero).
The Dirac spinors serve in (1) as basis vectors of the representation of the 
manifold fundamental group, while the electromagnetic field components prove to be
components of a curvature tensor of the manifold-covering space, and 
4-potentials serve as connections in this space. Energy,
momentum components, mass, spin and particle-antiparticle states
appear to be geometrical characteristics of the above manifold.
\par\medskip
\noindent{\bf Geometrization of weak interaction}
\par\medskip
Does the above topological interpretation reflect any physical reality? It is 
impossible to give an answer within the framework of the 
considered one-particle low-energy approximation alone. 
New approach did not lead to predictions of new
electrodynamical effects and did not explain anything that had not been 
explained 
before (as, e.g., deviations from the Newtonian gravitational law that was 
predicted 
by the theory of general relativity). So, we have to apply new concept to such
problems where we encounter some difficulties and where there is a 
possibility of 
obtaining new physical results. Within the one-particle approximation, 
these may be
some difficulties of the standard model of weak and strong interactions:
gauge invariance and P-T violation mechanisms, nucleon-nucleon
interaction at law energies, and so on. For many-particle quantum physics
these may be problems of quantum chemistry and problems of quantum field
theory. So, we have to choose the next step, and we will, at first, attempt
to apply topological approach to consider weak interaction within 
one-particle approximation, where we need not solve the problem of
geometrization of the second quantization procedure. In this section, we 
show that the geometrical concept leads to some new results, as  
compared with standard model.

As for the electromagnetic interaction, we suppose that the 
neutrino field together with its
weak field can be considered as another kind of space-time
4-manifold. We have shown in our previous publications (see [1-3])
that topological properties of the manifold
were represented by geometrical and symmetry properties of this 
manifold-covering space (the space with semimetrical translation in the case of 
interacting electron-positron and electromagnetic fields). Therefore, 
we have to
find out the type of new manifold-covering space that could give an
opportunity to reflect the known features of weak interaction.

     Let us now show that, in the one-particle approximation 
adopted in this work (low energies), weak interaction can be 
represented as a manifestation of the torsion in the covering 
space of a 4-manifold representing weak field and its sources.
Note that in due time, Einstein attempted at including electromagnetic 
field into a unified geometrical description of physical fields 
by "adding" torsion to the Riemannian space--time curvature, 
which reflects the presence of a gravitational field in general 
relativity [7]. Since the curvature of covering space 
corresponds, within our approach, to the electromagnetic field, we will 
attempt to include weak interaction into the topological approach by 
including torsion in this space.

It is known that weak interaction breaks the mirror space-time symmetry. On
the other hand this symmetry can be violated in space with torsion
(at least, left screw looks like a right one in the mirror). So,
it is natural to assume that, within our topological approach, 
torsion may be due to weak interaction. Let us 
consider the case where the electromagnetic field is 
absent, i.e., the curvature of covering space is zero.
     A space with torsion but without curvature is so called "the 
space with absolute parallelism" [7,8]. Thus, at first the challenge is to 
determine how does Eq.(1) change if the 
interparticle interaction is due only to torsion, which 
transforms the pseudo-Euclidean covering space into a space with 
absolute parallelism. 

Let us denote the torsion tensor by $S^k_{lm}$; then 
the problem can be formulated as follows. It is necessary to 
"insert" the tensor $S^k_{lm}$ or some of its components into Eq. (1) 
(where now $A_l=0$) so 
that the resulting equation remains invariant about the Lorentz 
transformations and adequately describes the experimental data 
(e.g., violation of spatial and time symmetry by weak 
interaction).
     Among the spaces with torsion, there are so-called "spaces 
with semisymmetric parallel translation" [6,9]. The torsion tensor 
$S^k_{lm}$ for such spaces is defined by the antisymmetric part of 
connection and can be represented as [6]
$$S^k_{lm}=K_lI^k_m-S_mA^k_l.\eqno (4)$$
Here, $K_l$ is a 4-vector and $I^k_l$ is the identity tensor. We will
investigate the above-mentioned kind of spaces as a 
candidate for the covering space because
this space has some preferential direction (vector $K_l$) and, so, may in
a way, be "responsible" for the P and T violation.

The vector $K_l$ has the property that an 
infinitesimal parallelogram remains closed upon the parallel 
translation in the hyperplanes perpendicular to this vector [6,9]. This 
indicates that, in the considered space, the sliding symmetry of Eq.(1) 
(the product of reflection operators $\gamma_l$ and translation operators 
$\partial/\partial x_l$) can only be retained in the 3-hyperplanes
perpendicular to $K_l$, while the translation symmetry can be
retained along $K_l$. This is the first thing we have to take into account
when looking for the analogy of Eq.(1) to describe a 
neurino and its weak field.

The second important point is that, unlike the case of 
electromagnetic interaction, we can now fix the invariant orientation 
in every point of the cavering space. It follows from the fact that,
in addition to vector $K_l$, we may introduce into consideration one more
vector $B_l$
$$B_l=S^p_{lp}=K_lA^p_p-K_pA^p_l.\eqno (5)$$
Having two nonparallel vectors $K_l$ and $B_l$ we can define the antisymmetric
second-rank tensor (bivector) $T_{lm}$
$$T_{lm}=\frac {1}{2}(K_lB_m-B_lK_m),\eqno (6)$$
and it is known that every bivector defines orientation in the space [8,9].

After taking into account the above symmetry properties, 
we obtain the following 
analogy of Dirac Eq.(1) for a neutrino interacting with its weak
field
$$i\sigma_1\bigtriangledown_1\varphi-\sum_{\alpha=2}^4i\sigma_\alpha
\bigtriangledown_\alpha
\varphi+\frac {g^2}{2}T_{lm}\sigma_m\bigtriangledown_l\varphi=m_N\varphi,
\eqno (7)$$
where $m_N$ is the neutrino mass, $\sigma_1$ is a two-row unit matrix, 
$\sigma_{2,3,4}$ are the Pauli 
matices ($\sigma_{x,y,z}$), $\varphi$ is a two-component spinor, 
and $\bigtriangledown_l$ are the covariant derivatives
$$\bigtriangledown_l = \frac {\partial}{\partial x_l} - igD_l.\eqno (8)$$
Here, $igD_l$ is a connection in the considered covering space.
Note that we introduce the coupling
constant $g$ by inserting $B\to igB,\quad K\to igK,\quad D\to igD$. 

We see that
the last term on the left-hand side of Eq.(7) is of the second order 
in the small
coupling constant $g$, but this term plays the main role in the space-time 
symmetry violation.
The above term corresponds within the 
Lagrangian formalism to the new additional
pseudoscalar term $L_{ad}$ (in the standard Lagrangian)
$$L_{ad}=g^2T_{lm}\varphi^+\sigma_m\bigtriangledown_l\varphi.\eqno (9)$$                                    

The fields $D_l$ in Eqs.(7,8) can be obtained using the symmetry properties
of the covering space with semisymmetric translation. This symmetry must be 
the same after rotations in the 3-hyperplane perpendicular to the
vector $K_l$. Let us assume that $0X_1$ axis is aligned with the vector
$K_l$ after a certain rotation expressed 
by the $U(1)$ representation (note that this 
rotation
reflects the gauge invariance of electromagnetic interaction which we did not
include into consideration).
Then the above hyperplane is a usual 3-space and for the two-component spinors 
the above rotations in this space are represented by the
matrices of the $SU(2)$ representation. So, this symmetry proves to be 
just the $SU(2)$-gauge
invariance used in the standard model of weak interaction for calculating
weak fields ($D_l$ fields in our case). 
This means that the fields $D_l$ in Egs.(7,8) can be expressed
in terms of twelve Yang-Mills fields $D_l^{\alpha}$ ($\alpha=x,y,z$) [10]
$$D_l=\frac {1}{2}(D_l^x\sigma_x+D_l^y\sigma_y+D_l^z\sigma_z).\eqno (10)$$
Then, the weak field strength has the form
$$G_{lm}=\frac {\partial D_l}{\partial x_m}-\frac {\partial D_m}{\partial x_l}-
ig(D_lD_m-D_mD_l).\eqno (11)$$

Finally, the equations for weak interactions read
\par\medskip
$$i\sigma_1\bigtriangledown_1\varphi-\sum_{\alpha=2}^4i\sigma_\alpha
\bigtriangledown_\alpha
\varphi+\frac {g^2}{2}T_{lm}\sigma_m\bigtriangledown_l\varphi=m_N\varphi,
\eqno (12)$$
$$G_{lm}=\frac {\partial D_l}{\partial x_m}-\frac {\partial D_m}{\partial x_l}-
ig(D_lD_m-D_mD_l).\eqno (13)$$
$$\sum_{i=1}^4 \frac{\partial G_{ik}}{\partial x_i}=j_k^N,\quad
j_k^N=g\varphi*\sigma_k\varphi.\eqno (14)$$ 
$$D_l=\frac {1}{2}(D_l^x\sigma_x+D_l^y\sigma_y+D_l^z\sigma_z).\eqno (15)$$

In the following publications we will represent the interdependence between
$T_{lm}$ and $D_l$ (it is a purely geometrical problem).
\par\medskip
\noindent{\bf Gravitational interaction}
\par\medskip

Gravitational field in general relativity is geometrized by representing
classical trajectories of macroscopic bodies as geodesics in the curved 
Riemannian space-time. The field sources are also supposed to be 
macroscopic objects. In this paper we propose the 
hypothesis for the geometrisation of
gravitational field generated by microscopic sources of gravitational field 
(elementary gravitational charges).

The main idea is as follows. Electrons are microscopic elementary
sources of elecgtromagnetic field and both of them (field and sources)
are represented, within our topological approach, by the special 4-manifold. 
The topological properties of this manifold are described by the Dirac-Maxwell
equations (1-3). Neutrinos are microscopic elementary sources of weak field,
and both of them (field and sources) are represented by a different special
4-manifold and the topological properties of this manifold are described by
equations (7-11). It is natural to assume that elementary sources of
gravitational field are also certain microscopic particles and that both of
them
(field and these sources) can be considered as a special 4-manifold
and can be described by group-theoretical relations similar to
Eqs.(1-3) and Eqs.(12-15).

We obtained above relations for gravitational interaction assuming that
the covering space for corresponding 4-manifold is the Riemann space (the
same as in general relativity). Instead of the spinor fields describing
electromagnetic and weak interactions, we use vector field $\phi_l$.
Then the analogy of the Dirac equation for the free electron-positron 
field will be the known Lorentz-invariant Proca 
equation for vector field $\phi_l$ describing free particles with mass $m$
and spin $1$ [11]
$$\frac {\partial \phi_{lm}}{\partial x_m}=m^2\phi_l,\quad
\phi_{lm}=\frac {\partial \phi_l}{\partial x_m}-\frac {\partial
\phi_m}{\partial x_l}.\eqno (16)$$
We take the gravitational interaction into account by replacing 
usual derivatives in Eq.(16) by the covariant derivatives with the standard 
connection in Riemann space [8,9]. The resulting equations for the field 
sources are
$$\frac {\partial \phi_{lm}}{\partial x_m}-\Gamma^p_{ml}\phi_{pm}-
\Gamma^p_{mm}\phi_{lp}=m_G^2\phi_l,\eqno (17)$$
where the connection $\Gamma^p_{ml}$ has the standard form [8,9]
$$\Gamma^l_{mn}=\frac {1}{2}G^{lp}\left (\frac {\partial G_{pm}}
{\partial x_n}+\frac {\partial G_{pn}}{\partial x_m}-\frac {\partial G_{mn}}
{\partial x_p}\right).\eqno (18)$$
Here $G_{mn}$ is the metrics of Riemann space, $m_G$ is the
elementary gravitational charge.

The connection $\Gamma^l_{mn}$ is not a tensor with respect to the arbitrary
coordinate transformation [8,9]. This means that the physical 
observable variables
should be connected directly not with $\Gamma^l_{mn}$ but with some tensor
that is expressed through $\Gamma^l_{mn}$. As in general relativity, we use
the Ricci tensor $R_{ik}$ [8,9]
$$R_{ik}=R^p_{pi,k},$$
where $R^p_{si,k}$ is the Riemannian space curvature tensor. The Ricci tensor
may be expressed through $\Gamma^l_{mn}$ as [8,9]
$$R_{ik}=\frac {\partial \Gamma^l_{il}}{\partial x_k}-
\frac {\partial \Gamma^l_{ik}}{\partial x_l}-\Gamma^l_{ik}\Gamma^p_{lp}+
\Gamma^l_{ip}\Gamma^p_{kl}. \eqno (19)$$
These equations can be considered as an analogy of the first pair of the
Maxwell equations.

We suppose that the microscopic source of gravitational field is a
gravitational current $j_l^G$ of our vector field $\phi_l$ which has the
standard form [11]
$$j_l^G=-im_G\left(\phi^p* \frac {\partial \phi_p}{\partial x_l}-
\phi^p \frac {\partial \phi_p*}{\partial x_l}\right),\eqno (20)$$
Then the analog of the second pair of Maxwell equations will be
equation for the Ricci tensor
$$\frac{\partial R_{ik}}{\partial x_i}=j_k^G.\eqno (21)$$

Finally, we have the following system of equations describing, within
the framework of topological approach, 
a gravitational field generated by its microscopic sources,
$$\frac {\partial \phi_{lm}}{\partial x_m}-\Gamma^p_{ml}\phi_{pm}-
\Gamma^p_{mm}\phi_{lp}=m^2_G\phi_l,\eqno (22)$$
$$\phi_{lm}=\frac {\partial \phi_l}{\partial x_m}-\frac {\partial
\phi_m}{\partial x_l}.\eqno (23)$$
$$R_{ik}=\frac {\partial \Gamma^l_{il}}{\partial x_k}-
\frac {\partial \Gamma^l_{ik}}{\partial x_l}-\Gamma^l_{ik}\Gamma^p_{lp}+
\Gamma^l_{ip}\Gamma^p_{kl}. \eqno (24)$$
$$\frac{\partial R_{ik}}{\partial x_i}=
-im_G\left(\phi^p* \frac {\partial \phi_p}{\partial x_k}-
\phi^p \frac {\partial \phi_p*}{\partial x_k}\right).\eqno (25)$$

The proposed elementary gravitational charge $m_G$ (mass $m_G$ 
is not yet defined),
represents one more candidate for a rather numerous family of 
so-called WINP particles (Weakly Interacting Neutral Particle [12]).
The possibilities of experimental confirmation will be
considered in detail elsewhere. Note in conclusion that, contrary to
general relativity, the Riemann space appears here not as a real curved
space-time but as an "effective" Riemann space (covering space used for 
the formal discription of a
more complicated geometrical object---microscopic nonmetrized closed 
topological space-time 4-manifold).
\par\medskip

\noindent{\bf Conclusion}
\par\medskip
We showed that, as in the case of electromagnetic interaction, the 
topological concept of quantum mechanics also does not contradict the known 
equations for weak interaction, although leads to some new small terms
that are responsible for the P-T violation.

Within the new approach the $U(1)SU(2)$ gauge invariance 
seems to be the standard Lorentz
rotation to new frame position, where electromagnetic and weak fields
are most symmetric.

The hyposesis is also proposed about new elementary particle that can be
considered as an elementary microscopic gravitational charge, and the equations
discribing these particles and their gravitational fields are suggested.

\par\medskip
\noindent
{\bf References}
\par\medskip
\noindent 
1. J. D. Bjorken, S. D. Drell, Relativistic quantum mechanics, McGraw-Hill Book
Company, 1964\\
2. O. A. Olkhov, Proc.7th Int.Symp. on Particles, Strings and
Cosmology, Lake Tahoe, California, 10-16 December 1999. Singapure-New Jersey-
Hong-Kong, World Scientific, 2000, p.160, quant-ph/0101137\\
3. O. A. Olkhov, Proc.XXIV Workshop on High Energy Physics and
Field Theory, Protvino, 27-29 June 2001, Protvino 2001, p.327, e-print 
hep-th/0205120;\quad  Proc.Int.Conf.on Structure and Interactions of the Photon,
Ascona, Switzerland, 2-7 September 2001. New Jersey-London-Singapure-Hong Kong,
World Scientific, 2002, p.360;\quad Proc.Int.Conf.on New Trends in
High-Energy Physics, Yalta (Crimea), 22-29 September, 2001, p.324\\
4. Ž. €. Olkhov, Chimitcheskaya fizika, 21,N1(2002)49; e-print hep-th/0201020\\
5. O. A. Olkhov, Preprint of Moscow Institute of Physics and Technology,
Moscow, 2002\\
6. J. A. Schouten and D. J. Struik, Einf\"uhrung in die neueren Methoden der
Differentialgeometrie, I. Groningen--Batavia.: Noordhoff, 1935\\
7. A. Einstein, Riemann Geometrie mit Aufrechterhaltung des 
Fernparallelismus.
Sitzungsber: preuss. Akad. Wiss., phys-math. K1., 1928, 217-221\\
8. P. K. Raschevski, Rimanova geometriay i tensornii analiz, Ch.3, Moscow,
Izdatelstvo techniko-teotetitcheskoi literaturi, 1953\\
9. J. A. Schouten, Tensor analysis for physicists. Oxford, 1952\\
10. C. N. Yang, R. L. Mills, Phys. Rev 96(1954)191\\
11. A. I. Achiezer, S. V. Peletminski, Fields and fundamental interactions,
Kiev, 1986\\
12. H. V. Klapdor-Kleingrothaus and K. Zuber, Teilchenastrophysik, 
Stuttgart, 1997

\end{document}